\documentclass[twocolumn,showpacs,preprintnumbers,amsmath,amssymb]{revtex4}

\usepackage{graphicx}
\usepackage{dcolumn}
\usepackage{bm}

\newcommand{\bra}[1]{\langle#1\rvert}
\newcommand{\ket}[1]{\lvert#1\rangle}

\renewcommand{\vec}[1]{\bm{\mathrm{#1}}}

\expandafter\ifx\csname
 natexlab\endcsname\relax\def\natexlab#1{#1}\fi
 \expandafter\ifx\csname bibnamefont\endcsname\relax
   \def\bibnamefont#1{#1}\fi
 \expandafter\ifx\csname bibfnamefont\endcsname\relax
   \def\bibfnamefont#1{#1}\fi
 \expandafter\ifx\csname citenamefont\endcsname\relax
   \def\citenamefont#1{#1}\fi
 \expandafter\ifx\csname url\endcsname\relax
   \def\url#1{\texttt{#1}}\fi
 \expandafter\ifx\csname
 urlprefix\endcsname\relax\def\urlprefix{URL }\fi
 \providecommand{\bibinfo}[2]{#2}
 \providecommand{\eprint}[2][]{\url{#2}}

\begin{document}

\title{Quantum memory for light using extended atomic ensembles in a tunable cavity}

\author{Alexey Kalachev}
\email{kalachev@kfti.knc.ru} \affiliation{Zavoisky
Physical-Technical Institute of the Russian Academy of Sciences,
Sibirsky Trakt 10/7, Kazan, 420029, Russia
}%

\date{\today}

\begin{abstract}
Cavity-assisted storage and retrieval of single-photon wave packets in optically thin spatially extended resonant materials are analyzed. It is shown that the use of cavity tuning allows one to store and recall time-symmetric double-sided exponential pulses with near unit efficiency. The optimal regime of the cavity tuning is determined and the effect of time jitter on the storage efficiency is analyzed.
\end{abstract}


\maketitle

\section{\label{sec:level1}Introduction}
A quantum memory for light may be defined as a controllable delay line for quantum states of the electromagnetic field. Such devices are currently
considered as a basic ingredient for scalable all-optical quantum computers \cite{KMNRDM_2007} and quantum repeaters \cite{BDCZ_1998}, which makes
them a topic of active research.  Although the optical quantum storage can be realized by means of usual delay lines \cite{PF_2002,LR_2006}, the most
promising approaches to the problem involve interaction of light with an atomic system controlled by additional fields or pulses
\cite{KM_1993,CZKM_1997,FYL_2000,FL_2000,KMP_2000,MK_2001,JSCFP_2004,KK_2006,NWRSWWJ_2007}. An important class of optical quantum memories is
designed for single-photon quantum states which can be stored and recall using interaction with a single atom \cite{CZKM_1997} and with an ensemble
of atoms in the regime of electromagnetically-induced transparency \cite{FYL_2000,FL_2000,CMJLKK_2005,DAMFZL_2005}, stimulated Raman absorption
\cite{NWRSWWJ_2007} and photon echo \cite{KM_1993,MK_2001,NK_2005,KTG_2005,ALSM_2006,ALSM_2007,HLALS_2008,ASRG_2008,HHSOLB_2008}.
Regardless of the approach, the possibility of manipulating the atomic system strongly depends on its physical properties, which, generally speaking,
restricts the set of materials suited for quantum storage. For example, the photon echo generated using controlled reversible inhomogeneous
broadening in a solid-state material \cite{NK_2005,KTG_2005,ALSM_2006,ALSM_2007,HLALS_2008} is of particular interest since it allows storage and
retrieval without application of additional laser pulses. But the frequency manipulation via the linear Stark effect is possible only if an optical
center occupies a non-centrosymmetric site in a medium (see the review in \cite{M_2007}). Besides, a significant difference between the excited- and
ground-state dipole moments is desirable. In this respect the scheme for quantum storage on subradiant states in a solid-state material
\cite{KK_2006} may be of interest. The advantage of the scheme is that storage and retrieval of a single-photon state may in principle be implemented
without any additional control fields or pulses acting on the resonant atomic system, which makes it possible to use materials that are difficult to
control.  In the present work we follow this approach and consider the possibility of such a passive manipulation of atomic states in a tunable
cavity. Enclosing an atomic ensemble in a cavity makes it possible to achieve high efficiency of quantum storage with optically thin materials,
thereby reducing phase relaxation rates and parasitic absorption, to the extend possible.

Atom-cavity systems have been investigated in the context of quantum memory in a series of papers
\cite{CZKM_1997,FYL_2000,DP_2004,DBP_2005,DCPG_2006,RS_2007,GALS_2007_1,HRGCD_2008}. In particular, it was shown that time-symmetric single-photon
wave packets can be stored and recalled with nearly unit probability, provided that atom-cavity interaction is controlled by special laser pulses
\cite{CZKM_1997}. Furthermore, the storage and retrieval of photons in a $\Lambda$-type atomic medium enclosed in a running-wave cavity were
optimized for different approaches to quantum storage \cite{GALS_2007_1}. It was shown that the optimal storage process is the time-reverse of
retrieval and the highest possible total efficiency of a quantum memory device is equal to $(C/(1+C))^2$, where $C$ is the cooperativity parameter.
The same efficiency can be achieved for the squeezing transfer between an atomic ensemble and fields in an optical cavity \cite{DP_2004}. Besides,
the storage efficiency was  analytically studied for two-level and various three-level atomic configurations and for various single-photon pulse
shapes \cite{RS_2007}. In the present work the capabilities of the storage and retrieval via cavity tuning are considered. The simplest strategy for
light storage and retrieval involves steplike modulation of the atom-field coupling strength via cavity tuning \cite{HRGCD_2008}: the coupling is switched off after absorption of a light pulse and switched on before the emission. In this case, the optimal pulse shape leading to the highest possible efficiency is asymmetric one and the output pulse is the time-reversed replica of the input pulse \cite{GALS_2007_1,HRGCD_2008}.
In this paper, a simple scheme for manipulation of cavity parameters is developed such that symmetric double-sided exponential pulses can be stored
and recall with a near maximum efficiency. The cavity tuning is apparently the counterpart of Stark shifting in an atomic system. In this connection
it should be noted that in Ref.~\cite{FRDM_2007} Stark shifting in an atomlike system was used to control its interaction with emf in a cavity and
possibilities of pulse shaping during spontaneous emission were investigated by quantum trajectory simulations. In comparison with that work, we
consider a full storage cycle --- i.e., the storage followed by retrieval --- and the optimal cavity control during the storage and retrieval of
information is described analytically without the use of adaptive algorithms.

The paper is organized as follows. In Sec.~II, we introduce a physical model of cavity-assisted interaction between an extended atomic ensemble and a
single-photon wave packet and present a general solution of the evolution equations in the case of dynamic cavity tuning. In Sec.~III, the optimal
regime of the cavity tuning that allows one to store and recall double-sided exponential light pulses is described analytically, the efficiency of
the storage followed by retrieval is calculated and the effect of time jitter on the storage efficiency is analyzed.

\section{\label{sec:level1}The model and basic equations}

Consider a system of $N\gg 1$
identical two-level atoms which are placed in a single-ended ring cavity and interact among themselves and with the external world only through the
electromagnetic field (Fig.~1). We are interested in the interaction of the atomic system with a resonant single-photon wave packet propagating through the cavity. In doing so we suppose that the bandwidth of collective spontaneous emission in the cavity as well as that of the incoming light are much smaller than the cavity linewidth. Such a cavity-assisted interaction corresponds to the strong "bad cavity" limit \cite{RC_1988} and can be considered in the Born-Markov approximation. Besides, we restrict the consideration to a single cavity mode which subtends a small solid angle at the atoms, $\alpha_\text{cav}$, so that spontaneous emission rate into free-space modes is actually not modified by the cavity. On the other hand, $\alpha_\text{cav}$ is supposed to be larger than the diffraction angle of collective atomic emission so that the cavity may be called "bad" in the angular domain as well.

We leave out any internal losses inside the cavity and consider only those through the partially transmitting mirror. The corresponding input-output relations for the field creation operators, provided that the mode wave-vector $\vec{k}$ is in $\alpha_\text{cav}$, may be written as
\begin{align}
\tilde{a}^\dag_\text{cav}&=\frac{it}{1-g(kp)}\,e^{-ikp/2}a^\dag_\text{in}
\equiv\sqrt{\mathcal{L}(kp)}\,e^{i\beta(kp)/2}a^\dag_\text{in},\\
a^\dag_\text{out}&={it}\,e^{-ikp/2}\tilde{a}^\dag_\text{cav}=e^{i\beta(kp)}a^\dag_\text{in},
\end{align}
where $g(kp)=r\exp(-ikp)$, $r=\sqrt{1-t^2}$, $t$ is the amplitude transmission coefficient of the semitransparent mirror and $p$ is the optical path length for one round trip through the cavity, which takes into account
all the refractive indices inside the cavity. The operator $\tilde{a}^\dag_\text{cav}$ is not normalized as a creation operator, but includes the enhancement of the cavity field and its phase shift with respect to the input field. As a result, the relations (1) and (2) have the same form as those for the complex field amplitude (see, e.g., Refs. \cite{S_1986,BR_2004} for details). Here $\tilde{a}^\dag_\text{cav}$ is defined for the central point of the atomic ensemble which is separated from the coupling mirror by the distance $p/2$ so that
\begin{gather}
\mathcal{L}(kp)=\frac{2\mathcal{F}}{\pi}\left(1+\left(\frac{2\mathcal{F}}{\pi}\right)^2\sin^2(kp/2) \right)^{-1},\label{L}\\
\beta(kp)=\arctan\left(\frac{(r^2-1)\sin(kp)}{2r-(1+r^2)\cos(kp)}\right),\label{beta}
\end{gather}
where $\mathcal{F}=\pi\sqrt{r}/(1-r)$ is the cavity finesse and it is assumed that $\mathcal{F}\gg 1$.

In the "bad cavity" limit, the values of the functions $\mathcal{L}(kp)$ and $\beta(kp)$ may be considered as constants within the atomic resonant line, which vary in accordance with Eqs.~(\ref{L}) and (\ref{beta}) under changes of $p$ in time. It is convenient to introduce the notations $\mathcal{L}(t)\equiv\mathcal{L}(kp(t))$ and $\beta(t)\equiv\beta(kp(t))$. The value of $\mathcal{L}(t)$ determines the atom-field coupling strength, while that of $\beta(t)$ describes phase shifts between the input and output fields. Therefore, in order to simplify the interaction and avoid unnecessary phase modulation of input and output photons, we put phase modulators PM3 and PM4 at the input and output of the cavity, respectively, each create phase shift $\exp(-i\beta(t)/2)$ upon the symmetrical modulation of the cavity optical length by modulators PM1 and PM2. As a result, the input-output relations become
\begin{gather}
\tilde{a}^\dag_\text{cav}=\sqrt{\mathcal{L}(t)}a^\dag_\text{in},\\
a^\dag_\text{out}=a^\dag_\text{in}\equiv a^\dag.
\end{gather}

\begin{figure}
\includegraphics[width=8.6cm]{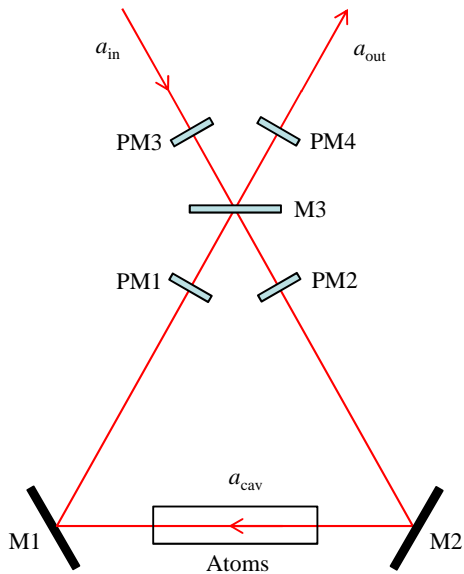}
\caption{\label{fig:levels} (Color online) Schematic of a quantum memory device with a tunable cavity. M1 and M2 are perfectly reflecting mirrors, M3 is a partially transmitting mirror. The optical path length of the cavity is controlled symmetrically by phase modulators PM1 and PM2, while modulators PM3 and PM4 are used for controlling input-output phase relations. All phase shifts may be implemented by movable mirrors thereby reducing the losses.}
\end{figure}

Let us denote the ground and excited states of $j$th atom ($j=1,\ldots,N$) by
$\ket{0_j}$ and $\ket{1_j}$, its position by $\vec{r}_j$
and resonance frequency by $\omega_0$. We assume that (i) the atoms are not moving as, for example, impurities embedded in a solid state material, (ii) the excitation volume may be approximated by a cylinder with the
cross section $S$ and the length $L$, and the Fresnel number of the
excitation volume $F=S(L\lambda)^{-1}\geq 1$, (iii) the atomic system has a small optical thickness,
$\alpha L \ll 1$, where $\alpha$ is a resonant absorption
coefficient. In the "bad cavity" limit, the Hamiltonian of the system, in the
interaction picture and rotating-wave approximation, is
\begin{equation}\label{Ham}
H=\sum_{j,\vec{k},s}\hbar g_{\vec{k},s}^\ast \sqrt{\mathcal{L}_{\vec{k}}(t)} b_j^\dag
a_{\vec{k},s}{\,e}^{i\vec{k}\cdot\vec{r}_j}
{\,e}^{i(\omega_0-\omega)t}+\text{H.c.}
\end{equation}
Here  $g_{\vec{k},s}=\frac{i}{\hbar}\left(\frac{\hbar\omega}{2\varepsilon_0
V}\right)^{1/2}(\vec{d}\cdot\vec{\varepsilon}_{\vec{k},s})$  is
the atom-field coupling constant, $b_j=\ket{0_j}\bra{1_j}$ is the
atomic transition operator, $a_{\vec{k},s}$ is the photon
annihilation operator in the radiation field mode with the
frequency $\omega=kc$ and polarization unit vector
$\vec{\varepsilon}_{\vec{k},s}$ ($s=1,2$), $V$ is the quantization
volume of the radiation field (we take $V$ much larger than the volume of the atom-cavity system), $\vec{d}$ is the dipole moment of
the atomic transition, and
\begin{equation}
\mathcal{L}_{\vec{k}}(t)=
\begin{cases}
\mathcal{L}(t),\quad \text{$\vec{k}$ in $\alpha_\text{cav}$},\\
1, \quad \text{otherwise}.
\end{cases}
\end{equation}
For simplicity we assume that
the vectors $\vec{\varepsilon}_{\vec{k},s}$ and $\vec{d}$ are
real.

We define the following collective atomic operators:
\begin{equation}
R=\sum_{j=1}^{N} b_j{\,e}^{- i\vec{k}_0\cdot\vec{r}_j},
\end{equation}
where $\vec{k}_0$ is directed along the cavity mode within the sample,
$|\vec{k}_0|=\omega_0/c$, so that the general form of the state of the
system can be written as
\begin{equation}\label{psi_0}
\ket{\psi(t)}=\sum_{\vec{k},s}f_{\vec{k},s}(t)\ket{0}\ket{1_{\vec{k},s}}+
P(t)\ket{1}\ket{\text{vac}}
\end{equation}
with normalization condition
$\sum_{\vec{k},s}|f_{\vec{k},s}(t)|^2+|P(t)|^2=1$,  where
$\ket{0}=\ket{0_1,0_2,\ldots,0_{N}}$ is the ground state of the
atomic system, $\ket{\text{vac}}$ is the vacuum state of
the radiation field,
$\ket{1_{\vec{k},s}}=a^\dag_{\vec{k},s}\ket{\text{vac}}$ and
$\ket{1}=N^{-1/2}R^{\dag}\ket{0}$.

Substituting Eqs.~(\ref{Ham}) and (\ref{psi_0}) in the
Schr\"{o}dinger equation we obtain
\begin{align}
\frac{\partial f_{\vec{k},s}(t)}{\partial
t}=&-ig_{\vec{k},s}\sqrt{N}\phi(\vec{k}_0-\vec{k})\sqrt{\mathcal{L}_{\vec{k}}(t)}
P(t){\,e}^{-i(\omega_0-\omega)t},\label{Eq1}\\
\frac{\partial P(t)}{\partial t}=&-i\sqrt{N}\nonumber\\
&\times\sum_{\vec{k},s}g^\ast_{\vec{k},s}\phi^\ast(\vec{k}_0-\vec{k})\sqrt{\mathcal{L}_{\vec{k}}(t)}
f_{\vec{k},s}(t){\,e}^{i(\omega_0-\omega)t},\label{Eq2}
\end{align}
where $\phi(\vec{x})=N^{-1}\sum_{j}{\exp}(i\vec{x}\cdot\vec{r}_j)$
is the diffraction function.

The photon density for the incoming wave packet at the input of the system reads
\begin{equation}
F_\text{in}(t)=\sqrt{\frac{Sc}{V}}\sum_{\vec{k},s}
f_{\vec{k},s}(-\infty){\,e}^{i(\omega_0-\omega)t},
\end{equation}
and for the emitted radiation we have the analogous equation with
$F_\text{in}(t)$ and $f_{\vec{k},s}(-\infty)$ replaced by
$F_\text{out}(t)$ and $f_{\vec{k},s}(t)$, respectively. Then the
solution of Eqs.~(\ref{Eq1}) and (\ref{Eq2}) may be written as (see Appendix)
\begin{equation}
F_\text{out}(t)=F_\text{in}(t)+\sqrt{2\gamma C(t)}\,P(t),\label{Solution_a}
\end{equation}
\begin{multline}
P(t)= P(-\infty)\exp\left(-\int_{-\infty}^t\gamma(C(\tau)+1)\,d\tau\right)\\
-\int_{-\infty}^t\sqrt{2\gamma C(\tau)}\,F_\text{in}(\tau)\exp\left(-\int_{\tau}^t\gamma(C(\tau')+1)\,d\tau'\right),\label{Solution_b}
\end{multline}
where $C(t)=\mathcal{L}(t)N\mu$ and $\gamma=1/T_2=1/2T_1$.
Here $\mu=3\lambda^2(8\pi S)^{-1}$ is a geometrical factor
\cite{RE_1971}, and
\begin{equation}\label{tau_R}
\frac{1}{T_1}=\frac{1}{4\pi\varepsilon_0}\frac{4d^2\omega_0^3}{3\hbar
c^3}
\end{equation}
is the Einstein A coefficient. The function $C(t)$, which can be expressed as $C(t)=\mathcal{L}(t)\alpha L/4$, is so-called cooperativity parameter and represents the effective optical depth of the medium in the cavity. The maximum value of the cooperativity parameter, $C_m=\alpha L \mathcal{F}/2\pi$, is attained at exact resonance and is equal to the optical length of the medium multiplied by the number of round trips during the photon life-time in the cavity. The strong "bad cavity" limit is now defined as $C_m\gamma\ll\Delta\omega_\text{cav}=2\pi c/\mathcal{F}p$.  According to the definition of $C(t)$ and Eqs.~(\ref{L}) and (\ref{beta}), there is a one-to-one correspondence between the values of $C(t)$ and those of $p(t)$ and $\beta(t)$ on each side of the cavity line. Thus, the cavity tuning is fully specified by the function $C(t)$.

\section{\label{sec:level1}Storage and retrieval of double-sided exponential pulses}

In this section we show how to tune the cavity length for the storage and retrieval of symmetric double-sided exponential pulses and calculate the efficiency of such a cavity-assisted quantum memory. There are several reasons for choosing such pulses for the analysis: first, symmetric pulses are suited for multiple storage and retrieval without resorting to additional time reversal; second, single-photon pulses of such a shape may be generated via spontaneous parametric down conversion when a nonlinear crystal is enclosed in a cavity and continuously pumped \cite{LO_2000,NNTVP_2007}; third, such pulses are near optimal for quantum-information processing from the viewpoint of tolerance against the effect of time jitter \cite{RRN_2005}.

To begin with, we determine the time dependence of the cooperative parameter $C(t)$ which leads to the spontaneous emission of the double-sided exponential pulses (Fig.~2). Suppose that the spontaneous emission begins at the moment of time $t=0$ when atoms are coherently excited so that $P(0)=1$. Let the maximum of the pulse amplitude falls at time $t=T$ when the cooperativity parameter reaches its maximum value $C_m$ by the cavity tuning and the latter does not change after that. In this case the output pulse can be written as
\begin{equation}\label{Fout}
F_\text{out}(t)=F_m \exp\left(-\gamma(C_{m}+1)|t-T|\right),\quad t\geq 0.
\end{equation}
The trailing part of the pulse ($t\geq T$) is the usual exponential decay when the cavity is tuned to resonance with the atomic transition. The leading part is the time-reversed exponential decay which the dependence $C(t)$ should be determined for. For this purpose, according to Eqs.~(\ref{Solution_a}) and (\ref{Solution_b}),
we need to solve the equation
\begin{multline}\label{Eq3}
\sqrt{2\gamma C(t)}\exp\left(-\int_0^t \gamma (C(\tau)+1)\,d\tau\right)\\=F_m\exp\left(\gamma(C_{m}+1)(t-T)\right),
\end{multline}
provided that $0\leq t\leq T$.
By differentiating the logarithm of Eq.~(\ref{Eq3}) we obtain Bernoulli equation for $C(t)$, which has the following solution:
\begin{equation}\label{C}
C(t)=\frac{C_{m}+2}{A(C_{m}+2)\,e^{-2\gamma(C_{m}+2)t}-1}.
\end{equation}
The constant $A$ is found from the condition $C(T)=C_{m}$ to be
\begin{equation}\label{A}
A=e^{2\gamma(C_{m}+2)T}\left(\frac{1}{C_{m}+2}+\frac{1}{C_{m}}\right).
\end{equation}
Finally, substituting Eqs.~(\ref{A}) and (\ref{C}) into the left side of Eq.~(\ref{Eq3}) we find
\begin{equation}
F_m=\left(
\frac{2\gamma}{\displaystyle e^{2\gamma T}\left(\frac{1}{C_{m}+2}+\frac{1}{C_{m}}\right)-\frac{e^{-2\gamma(C_{m}+1)T}}{C_{m}+2}}
\right)^{1/2}.
\end{equation}
Thus, if $T=0$, then $F_m=\sqrt{2\gamma C_{m}}$ and we have an ordinary exponential decay pulse with the highest possible amplitude. On the other hand, when $\gamma C_mT\gg 1$ then $F_m=  \sqrt{\gamma C_m(C_m+2)/(C_m+1)}\exp(-T/T_2)$ and we have a double-sided exponential pulse the maximum amplitude of which decreases with $T$ due to the phase relaxation caused by spontaneous emission into free-space modes. The norm of the output pulse is
\begin{align}\label{Norm}
\mathcal{N}&=\int_0^\infty F_\text{out}(\tau)^2\,d\tau\nonumber\\
&=F_m^2\left(\frac{2-e^{-2\gamma(C_m+1)T}}{2\gamma(C_m+1)}\right)\nonumber\\
&=\frac{2-e^{-2\gamma(C_m+1)T}}{e^{2\gamma T}\left(\displaystyle\frac{C_m+1}{C_m+2}+\frac{C_m+1}{C_m}\right)-\displaystyle\frac{C_m+1}{C_m+2}\,e^{-2\gamma(C_m+1)T}},
\end{align}
and to the lowest order on $1/C_m$ we have
\begin{equation}
\mathcal{N}=(1-C_m^{-2})\,e^{-2T/T_2}
\end{equation}
if $\gamma C_mT\gg 1$.

\begin{figure}
\includegraphics[width=7cm]{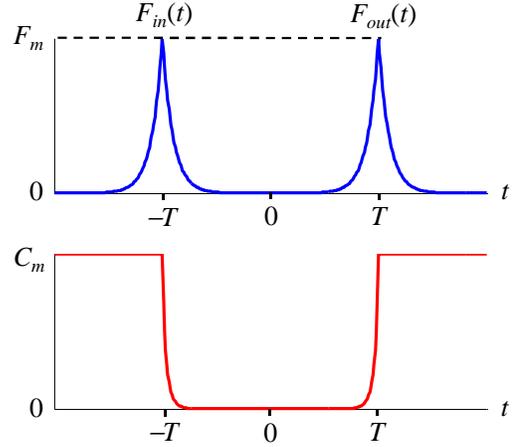}
\caption{\label{fig:levels} (Color online) The input, $F_\text{in}(t)$, and output, $F_\text{out}(t)$, double-sided exponential pulses (above) and corresponding cooperative parameter $C(t)$ (below) as functions of time.}
\end{figure}

It should be noted that the leading part of the output pulse can be generated without distortion only if the minimum value of $C(t)$ at the initial moment of time (far from resonance) is larger than the lowest possible value of the cooperative parameter, $C_m/\mathcal{F}^2$. It follows from Eqs.~(\ref{C}) and (\ref{A}) that in the case of $C_m\gg 1$ we need
\begin{equation}\label{Fidelity_min}
\mathcal{F}^2> 2\exp(2C_m\gamma T).
\end{equation}
This condition leads to extremely high values of the finesse for large delay times. Fortunately, as we shall see below, this is no problem in a real case of storage and retrieval during finite time intervals.

As noted in Introduction, the optimal strategy for maximizing the storage and retrieval efficiency in a quantum memory device involves the storage process that is the time-reverse of retrieval. Taking this into account we consider now the storage of the double-sided exponential single-photon pulse which is the time-reversed replica of the output pulse Eq.~(\ref{Fout}):
\begin{equation}\label{F_in}
F_\text{in}(t)=F_\text{out}(-t)/\sqrt{\mathcal{N}}.
\end{equation}
Thus, the input pulse begins at $t=-\infty$, terminates at $t=0$, and its maximum falls at $t=-T$. During the interaction between the input pulse and the atomic system the time-dependencies of $p(t)$, $\beta(t)$ and, consequently, $C(t)$ are assumed to be time-reversed replicas of those at the emission stage. In this case, from Eq.~(\ref{Solution_b}), assuming $P(-\infty)=0$, we obtain
\begin{align}
P(0)&=-\int_{-\infty}^0 F_\text{in}(\tau)F_\text{out}(-\tau)\,d\tau \nonumber\\
&=-\frac{1}{\sqrt{\mathcal{N}}}\int_{-\infty}^0 F_\text{out}(-\tau)^2\,d\tau =-\sqrt{\mathcal{N}}.
\end{align}
The total efficiency of a quantum memory device --- i.e., the efficiency of storage followed by retrieval --- may be defined as
\begin{equation}
\mathcal{E}=\frac{\int_0^\infty F^2_\text{out}(\tau)\,d\tau}{\int_{-\infty}^0 F^2_\text{in}(\tau)\,d\tau}.
\end{equation}
The denominator is equal to unity by definition (\ref{F_in}), whereas the numerator is equal to $P(0)^2\mathcal{N}$ so that we obtain
$\mathcal{E}=\mathcal{N}^2$.
In the limit of $C_m\gg 1$, the efficiency becomes $\mathcal{E}\approx \exp(-2T/T_2)$, i.e., the losses are entirely due to the phase relaxation on the resonant transition during the delay time between the input and output pulses, $2T$.

In a more realistic case, the processes of storage and retrieval are finite in time. Since the present scheme is aimed at the processing of symmetric pulses, we proceed by considering truncated storage and retrieval during the time windows $[-2T, 0]$ and $[0, 2T]$, respectively, provided that the input pulse has the same form as Eq.~(\ref{F_in}). In this case
\begin{align}\label{Norm2}
\mathcal{N}_\text{trunc}&=F_m^2\left(\frac{1-e^{-2\gamma(C_m+1)T}}{\gamma(C_m+1)}\right)\nonumber\\
&=\frac{2\left(1-e^{-2\gamma(C_m+1)T}\right)}{2-e^{-2\gamma(C_m+1)T}}\mathcal{N}.
\end{align}
Whereas the norm of the infinite output pulse, $\mathcal{N}$, decreases with $T$ monotonically, the norm of the truncated pulse, $\mathcal{N}_\text{trunc}$, which is equal to zero when $T=0$, has a maximum at some value of $T$. The reason is that the larger $T$, the smaller the maximum amplitude $F_m$, but, on the other hand, the smaller a truncated part of the pulse. It is follows from Eq.~(\ref{Norm2}) that in the limit of $C_m\gg 1$, the maximum value of the norm
\begin{equation}
(\mathcal{N}_\text{trunc})_\text{max}\approx 1-\frac{\ln(C_m/2)+1}{C_m}
\end{equation}
is achieved when the time window is
\begin{equation}\label{Tmax}
2T_\text{max}\approx\frac{1}{\gamma C_m}\ln\left(\frac{C_m}{2}\right).
\end{equation}
Thus, when the double-sided exponential pulses are absorbed and emitted during the optimal finite time intervals, the total efficiency of storage followed by retrieval becomes
\begin{equation}
\mathcal{E}=1-2/C_m-2\ln(C_m/2)/C_m,
\end{equation}
which is less than the highest possible value $(C_m/(1+C_m))^2\approx 1-2/C_m$. The difference is obviously due to the phase relaxation on the resonant transition, which in our model is entirely connected with spontaneous emission into free-space modes, during the storage and retrieval of information. For a given efficiency, the delay $2T_\text{max}$ can be considered as a minimum (intrinsic) storage time. Additional storage of a photon for some time, say $T_s$, simply introduces $e^{-T_s/T_2}$ decay into the amplitude of the output pulse.

In the case of the optimal time window, $2T_\text{max}$, the condition (\ref{Fidelity_min}) reduces to
$\mathcal{F}^2>C_m$ or $\mathcal{F}>\alpha L/2\pi$, which is always valid as optically thin materials are used.
On the other hand, if $\mathcal{F}\gg\alpha L/2\pi$, then only small deviations of $p$ from the exact resonance --- i.e., $\delta p/\lambda\sim \sqrt{C_m}/\mathcal{F}\ll 1$ --- are necessary for implementing storage and retrieval.

An important question is the influence of time jitter --- i.e., the time uncertainty
of receiving the input photon within the fixed storage window --- on the efficiency. If we assume that the maximum of the double-sided exponential input pulse (\ref{F_in}) falls at the time $-T+\delta$, then the amplitude of the excited atomic state created at the end of the storage process, $P(0)$, scales as
\begin{equation}
P(0)\propto\int_{0}^{2T} e^{-\gamma(C_m+1)|\tau-T|}\,e^{-\gamma(C_m+1)|\tau-T-\delta|}\,d\tau.
\end{equation}
Taking the optimal storage window, $T=T_\text{max}$, we obtain
\begin{equation}
\frac{P_{\delta\neq 0}(0)}{P_{\delta=0}(0)}=\frac{1+\gamma C_m|\delta|-\left(1+e^{2\gamma C_m |\delta|}\right)/C_m}{1-2/C_m}e^{-\gamma C_m|\delta|}
\end{equation}
for $C_m\gg 1$ and $|\delta|<T$. According to this formula, the total efficiency, which scales as $({P_{\delta\neq 0}(0)}/{P_{\delta=0}(0)})^2$, is reduced by 10\% (1\%) when $\gamma\delta C_m\approx 0.35$ $(0.1)$ and $C_m>100$. In other words, quantum memory efficiency of 90\% (99\%) requires temporal
synchronization to within 35\% (10\%) of the pulsewidth. The same limitations arise when considering fidelity of quantum gates \cite{RRN_2005}.

Finally, we discuss possible experimental conditions under which the storage and retrieval of the double-sided exponential pulses is possible. As a promising material for the quantum storage we consider rare-earth-metal-doped crystals (see the review in \cite{M_2002}) in which phase relaxation time at liquid helium temperature can reach several ms \cite{STCEH_2002}.
Though the inhomogeneous broadening of resonant lines in such crystals are usually several GHz, a narrow absorption line on a nonabsorbing background may be prepared using hole-burning techniques \cite{PSM_2000,SPMK_2000,NROCK_2002,SLLG_2003,NRKKS_2004,RNKKS_2005}. The absorption linewidth is determined by the laser linewidth used for the preparation and is bounded from below only by the homogeneous broadening. Taking the absorption linewidth, $\Gamma=10$~kHz, we obtain $T_2=1/\pi\Gamma=32~\mu$s. The finesse of the cavity with an impurity crystal is limited mainly by absorption losses in the host material which are typically of the order of $10^{-2}\text{~cm}^{-1}$. So, taking a thin crystal, $L=1$~mm, it is possible to reach $\mathcal{F}\sim 10^3$ and obtain $C_m=100$ for $\alpha L\sim 10^{-1}$. In such a case, the duration of input and output pulses is equal to 320~ns, the optimal duration of storage and retrieval processes $2T_\text{max}\approx 1.25~\mu$s, and the overall efficiency $\mathcal{E}\approx 90$\%.

\section{\label{sec:level1}Conclusion}

It is shown that double-sided exponential single-photon wave packets resonant with a transition in an extended ensemble of atoms enclosed in a
tunable cavity may be stored and recall with a near unit probability using cavity tuning only. The developed scheme readily can be combined with that
for passive creation of orthogonal subradiant states considered in \cite{KK_2006}, which in principle allows one to organize multichannel quantum
storage on a homogeneously broadened resonant transition for time-symmetric pulses. Besides, storage and retrieval on several frequencies can also be
performed using periodic absorption structures proposed in \cite{ASRG_2008}, provided that the cavity linewidth is sufficiently large.

\begin{acknowledgements}
The author would like to thank Stefan Kr\"{o}ll for useful comments. This work was supported by the Program of the Presidium of RAS 'Quantum
macrophysics'.
\end{acknowledgements}

\appendix*
\section{}
In order to obtain the solution for $F(t)$ we integrate Eq.~(\ref{Eq1}), which gives
\begin{multline}\label{F}
f_{\vec{k},s}(t)=f_{\vec{k},s}(0)-ig_{\vec{k},s}\sqrt{N}
\phi(\vec{k}_0-\vec{k})\\
\times\int_0^t \sqrt{\mathcal{L}_{\vec{k}}(\tau)}P(\tau){\,e}^{-i(\omega_0-\omega)\tau}d\tau.
\end{multline}
Then we substitute this result into the general formula
\begin{equation}
F(t)=\left(\frac{Sc}{V}\right)^{1/2}
\sum_{\vec{k},s}f_{\vec{k},s}(t){\,e}^{i(\omega_0-\omega)t},
\end{equation}
convert the sum to an integral,
\begin{equation}\label{SumToInt}
\sum_{\vec{k}}\to\frac{V}{(2\pi)^3c^3}\int_0^\infty\omega^2\,d\omega\int
d\Omega_{\vec{k}},
\end{equation}
and take into account that in the Born--Markov approximation,
\begin{equation}
\int_0^t
d\tau{\,e}^{i(\omega_0-\omega)\tau}=\pi\,\delta(\omega_0-\omega)
+i\frac{\mathcal{P}}{\omega_0-\omega},
\end{equation}
where the imaginary part may be neglected \cite{RadiativeCorrections}, and
\begin{multline}\label{mu1}
\int
d\Omega_{\vec{k}}\sum_s(\vec{d}\cdot\vec{\varepsilon}_{\vec{k},s})
\phi(\vec{k}_0-\vec{k})\sqrt{\mathcal{L}_{\vec{k}}(t)}\\=\frac{16\pi}{3}\mu d\sqrt{\mathcal{L}(t)},
\end{multline}
where $\mu$ is a geometrical factor \cite{RE_1971}. The latter is equal to $3\lambda^2(8\pi
S)^{-1}$ for a pencil-shaped cylindrical volume with the cross-section $S$, the Fresnel number $F\geq 1$, a
vector $\vec{k}_0$ lying along the axis of the cylinder and dipole
moments oriented perpendicular to the axis. In calculating (\ref{mu1}) we take into account that the diffraction angle of collective emission, which is given by the function $\phi(\vec{k}_0-\vec{k})$ and of the order of $4\pi\mu$, is smaller than $\alpha_\text{cav}$.
As a result, we obtain Eq.~(\ref{Solution_a}).

To obtain the solution for $P(t)$ we substitute the solution Eq.~(\ref{F})
into Eq.~(\ref{Eq2}), employ Eq.~(\ref{SumToInt}) and take into account that
\begin{multline}
\int
d\Omega_{\vec{k}}\sum_s(\vec{d}\cdot\vec{\varepsilon}_{\vec{k},s})^2
\phi^2(\vec{k}_0-\vec{k})\mathcal{L}_{\vec{k}}(t)\\=\frac{8\pi}{3}(\mu+{1}/{N})d^2\mathcal{L}(t).
\end{multline}
This leads to the following equation
\begin{equation}
\frac{\partial P(t)}{\partial t}=-\gamma(1+C(t))P(t)-\sqrt{2\gamma C(t)}F_\text{in}(t),
\end{equation}
which has a general solution of the form (\ref{Solution_b}).


\begin{thebibliography}{46}
\bibitem{RadiativeCorrections}
The Lamb shift may be neglected if the density of states of the cavity remains symmetric with respect to the atomic frequency $\omega_0$. With this
in mind, the atomic system should be placed in two cavities which sustain different spatial modes and undergo opposite frequency detunings.
\end{thebibliography}

\end{document}